\documentclass[12pt]{article}

\newcommand{\GE}{\Gamma_{11}}

\begin{document}
\vspace*{-.6in}
\thispagestyle{empty}
\begin{flushright}
CALT-68-2088\\
hep-th/9612080
\end{flushright}
\baselineskip = 20pt

\vspace{.5in}
{\Large
\begin{center}
Gauge-Invariant and Gauge-Fixed D-Brane Actions\footnote{Work supported in part by
the U.S. Dept. of Energy under Grant No. DE-FG03-92-ER40701.}
\end{center}}

\begin{center}
Mina Aganagic, Costin Popescu, and John H. Schwarz\\
\emph{California Institute of Technology, Pasadena, CA  91125 USA}
\end{center}
\vspace{1in}

\begin{center}
\textbf{Abstract}
\end{center}
\begin{quotation}
\noindent The first part of this paper presents
actions for Dirichlet $p$-branes embedded in a flat 10-dimensional
space-time. The fields of the ($p+1$)-dimensional world-volume theories 
are the 10d space-time coordinates $ X^m$, a pair of
Majorana--Weyl spinors $\theta_1$ and $\theta_2$, and a $U(1)$ gauge
field $A_{\mu}$. The N = 2A or 2B super-Poincar\'e group in ten
dimensions is realized as a global symmetry. 
In addition the theories have local symmetries consisting of general coordinate 
invariance of the world volume, a local fermionic symmetry (called ``kappa''), 
and  $U(1)$ gauge invariance. A detailed proof of the kappa symmetry is given
that applies to all cases ($p = 0,1, \ldots, 9$). The second part of the paper presents
gauge-fixed versions of these theories. The fields of the 10d ($p=9$) gauge-fixed
theory are
a single Majorana--Weyl spinor $\lambda$ and the $U(1)$ gauge field
$A_{\mu}$. This theory, whose action turns out to be surprisingly simple,  is
a supersymmetric extension of 10d Born--Infeld theory. 
It has two global supersymmetries:
one represents an unbroken symmetry, and the second corresponds to a broken
symmetry for which $\lambda$ is the Goldstone fermion. The gauge-fixed 
supersymmetric D-brane theories
with $p<9$ can be obtained from the 10d one by dimensional reduction.
\end{quotation}
\vfil

\newpage

\pagenumbering{arabic} 

\section{Introduction}

D-branes have become an active area of study during the 
past year~\cite{polchinski1,polchinski2}. Many of their
remarkable properties have been elucidated~\cite{witten,danielsson},
and the coupling of their bosonic degrees of
freedom to bosonic background fields
has been worked out~\cite{douglas,green2}.
In particular, they have provided a powerful
tool for the study of black holes in string theory~\cite{strominger}.  Very
recently an interesting proposal for understanding non-perturbative 11d (M
theory) physics in terms of ensembles of D 0-branes has been put
forward~\cite{banks}.  For these reasons and more it is desirable to achieve as
thorough an understanding of D-branes as possible.  

It has been known for some time that D-brane
world-volume theories contain a $U(1)$ gauge field, whose
self interactions are described by Born--Infeld theory~\cite{born}--\cite{bachas}.
However, only recently has attention turned to the study of supersymmetric
Born--Infeld theories appropriate to the description of 
D-branes~\cite{cederwall1}--\cite{bergshoeff1}. In these works the
crucial ingredient 
is a local fermionic symmetry of the world volume theory called ``kappa
symmetry.''  It was first identified by Siegel~\cite{siegel} for the
superparticle~\cite{brink}, and subsequently applied to the
superstring~\cite{green1}.  Next it was simplified (to eliminate an unnecessary
vector index) and applied to a super 3-brane in 6d~\cite{hughes}.
Then came the super 2-brane in eleven dimensions~\cite{bergshoeff2}, followed by
all super $p$-branes (without world-volume gauge fields)~\cite{achucarro1,duff2}.

The main distinction between D-branes and the previously studied super $p$-branes
is that the field content of the world-volume theory includes an abelian vector
gauge field $A_\mu$ in addition to the superspace coordinates $(X^m,\theta)$ of
the ambient $d$-dimensional space-time.  In the case of super $p$-branes whose only
degrees of freedom are $(X^m,\theta),$ $(p + 1)$-dimensional actions have been
formulated that have super-Poincar\'e symmetry in $d$ dimensions realized as a
global symmetry.  In addition they have world-volume general coordinate
invariance, which ensures that only the transverse components of $X^m$ are
physical, and a local fermionic kappa symmetry, which  effectively
eliminates half of the components of $\theta$.  This symmetry reflects the fact
that the presence of the brane breaks half of the supersymmetry in $d$ dimensions,
so that half of it is realized linearly and half of it nonlinearly in the
world-volume theory.  The physical fermions of the world-volume theory
are the Goldstone fermions associated to the broken supersymmetries.

One example of a kappa-symmetric 
D-brane action was derived prior to the recent works cited above.  
As noted in~\cite{duff},
the super 2-brane action in 11d can be converted to the D 2-brane in 10d
by performing a duality transformation in the world volume theory that
replaces one of the $X$ coordinates by a world-volume $U(1)$
gauge field.  This has been
worked out in detail by Townsend~\cite{townsend}, and some of his formulas have
given us guidance in figuring out the generalization 
to all $p$.  (See also~\cite{schmidhuber}.)
One lesson of that example is that the analysis is simplified considerably by
{\it not} introducing an auxiliary world-volume metric field in the formulas.
Auxiliary metric fields have been included in most studies of super $p$-branes, 
though this was not necessary.
If one attempted to incorporate them in the D-brane formulas, this would create
considerable algebraic complications. (For a discussion of this issue see
ref.~\cite{cederwall3}.)

This paper has two primary purposes. The first one -- the subject of section 3 --
is to present formulas for D-brane actions with local
kappa symmetry analogous to those of the super $p$-branes.  
These results, which were reported 
very succinctly in our first paper~\cite{aganagic},
are described here in greater detail. There are a number
of non-trivial identities that are required to establish kappa symmetry
and other features of the theory,
which were stated without proof in our first paper. In this paper we
give very detailed proofs of these identities in a series of Appendices.

Like our first paper, this one only considers D-branes embedded in a flat non-compactified
10d space-time. More specifically, we
assume a flat type IIA background when $p$ is even
and a flat type IIB background when $p$ is odd.  
References \cite{cederwall2} and \cite{bergshoeff1} have gone further
and included the coupling to arbitrary Type II supergravity backgrounds
satisfying the supergravity field equations.   Perhaps because of
the restriction to flat backgrounds, our proofs of 
the identities required to establish kappa symmetry 
appear to be much simpler than in those works.

The second purpose of our paper -- the subject of
section 4 -- is to choose a physical gauge in which
all remaining fields represent dynamical degrees of freedom
of the world-volume theory. Specifically, we choose a ``static gauge'' in
which $p+1$ of the space-time coordinates are identified with the
world-volume coordinates.
In such a gauge the remaining spatial coordinates can be
interpreted as $9-p$
scalar fields representing transverse excitations of the brane.
In fact, they are the Goldstone bosons associated to spontaneously
broken translational symmetries. The gauge field $A_\mu$ gives 
$p - 1$ physical degrees of freedom, 
so altogether there is a total of 8 bosonic modes.
(This counting doesn't apply to the $p=0$ case, which is somewhat special.)
The 32 $\theta$ coordinates are cut in half by kappa symmetry and in half again
by the equation of motion, so they give rise to 8 physical fermionic degrees of freedom.
The really crucial step is to make the right choice of which half of the
$\theta$ coordinates to set to zero in the static gauge. If one doesn't make a
convenient choice, the formulas can become unwieldy. 
For example, ref.~\cite{achucarro2} studied a 
gauge choice for a super 2-brane in four dimensions for which
the resulting gauge-fixed theory turned out
to be very complicated.

Fortunately, at least for D-branes in type II theories, there is a gauge choice
for the $\theta$ coordinates that leads to simple tractable formulas. Specifically,
we set one of the two Majorana--Weyl spinors ($\theta_1$ or $\theta_2$) equal
to zero. Then the fields of the remaining gauge-fixed theory are just the
gauge field $A_{\mu}$, the remaining $\theta$ which we rename $\lambda$,
and the $9-p$ scalar fields mentioned above. In particular, for $p=9$ there is
only $\lambda$ and $A_{\mu}$, and the theory is a supersymmetric extension
of 10d Born--Infeld theory, which can also be thought of as a self-interacting
extension of 10d super-Maxwell theory. The action for this theory is
surprisingly simple. Since we are proud of the result, we exhibit it here:
\begin{equation}
 - \int d^{10} \sigma \sqrt{- {\rm det}\, \big(
\eta_{\mu\nu} + F_{\mu\nu} - 2 \bar\lambda \Gamma_{\mu}
\partial_\nu \lambda + \bar \lambda \Gamma^\rho \partial_\mu \lambda
\bar\lambda \Gamma_\rho \partial_\nu \lambda \big)}.
\end{equation}
This theory is invariant under two supersymmetries, which are just the
original global supersymmetries we started with, supplemented with compensating
gauge transformations required to maintain the gauge. 
The explicit infinitesimal transformations are given in section 4.2. One of them, whose
infinitesimal parameter we call $\epsilon_1$, corresponds to the supersymmetry
that is spontaneously broken by the presence of the brane. It is realized non-linearly
in the Volkov--Akulov manner~\cite{volkov}. These transformations are
rather simple, and the associated invariance of the expression given above
is easy to verify. The second supersymmetry (with parameter $\epsilon_2$)
is much more complicated. Since the $\epsilon_1$ symmetry can be 
realized by a large class of formulas, and the $\epsilon_2$ symmetry is so subtle,
it would have been practically impossible to discover this action and
its symmetries by ``brute force.''
Starting from a gauge-invariant action made it possible. The unbroken supersymmetry
corresponds to a combined transformation with $\epsilon_1 = \epsilon_2$.
Section 4.3 demonstrates that all 
other gauge-fixed D-brane actions with $p<9$, which are also
maximally supersymmetric Born--Infeld theories, can be obtained from this
action by dimensional reduction. Ones with less supersymmetry can then
be obtained by making appropriate truncations.

There have been some previous studies of supersymmetric extensions of Born--Infeld 
theory. For example, the special case of N = 1 in 4d has been described in
a superfield formalism. This example was worked out first in ref.~\cite{cecotti}
and elaborated upon recently in ref.~\cite{bagger}.  The latter work emphasized
the fact that the theory has both an unbroken supersymmetry 
and a non-linearly realized broken one, as required for D-branes.
In their formalism the unbroken supersymmetry is manifest and the broken one
is complicated. The fact that the reverse is the case for our formula suggests
that it may be rather difficult to work out the precise relationship between the
two descriptions. The leading corrections to 10d (and 6d)
super-Maxwell (and super Yang--Mills) theory implied by the Born--Infeld
structure of the gauge fields were studied in ref.~\cite{bergshoeff4}. 
A field redefinition is required to
establish the correspondence with the results obtained there.
Ref.~\cite{metsaev} studied fermionic terms in the superstring effective action
by comparing to the tree-level S matrix.

\medskip

\section{Preliminaries}

\subsection{Superspace Coordinates}

Our conventions are the following.  $X^m$, $m = 0, 1,\ldots, 9$, denotes the 10d
space-time coordinates and $\Gamma^m$ are $32\times 32$ Dirac matrices 
appropriate to 10d with
\begin{equation}
\{\Gamma^m, \Gamma^n\} = 2\eta^{mn}, \quad {\rm where}~ \eta = (- ++ \ldots +).
\end{equation}
These $\Gamma$'s differ by a factor of $i$ from those of ref.~\cite{green3}.
For this choice of gamma matrices the massive Dirac equation is 
$(\Gamma\cdot\partial - M)\Psi =0$.  In fact, our conventions are such that the
quantity $i=\sqrt{-1}$ will not appear in any equations. As is quite standard,
we also introduce $\Gamma_{11} = \Gamma_{0}\Gamma_{1}\ldots \Gamma_{9}$,
which satisfies $\{ \Gamma_{11}, \Gamma^m \} =0$ and 
$\big(\Gamma_{11}\big)^2 =1$.

The Grassmann coordinates $\theta$ are space-time spinors and world-volume
scalars.  When $p$ is even, which is the Type IIA case, 
$\theta$ is Majorana but not Weyl.  It can be
decomposed as
$\theta = \theta_1 + \theta_2$, where
\begin{equation}
\theta_1 = {1 \over 2} (1 + \Gamma_{11})\theta, \qquad \theta_2 = {1 \over 2}
 (1 - \Gamma_{11})\theta. \label{thetas}
\end{equation}
These are Majorana--Weyl spinors of opposite chirality.  When
$p$ is odd, which the Type IIB case,
there are two Majorana--Weyl spinors $\theta_\alpha$ $ (\alpha = 1,2)$
of the same chirality.  The index $\alpha$ will not be displayed explicitly.
The group that naturally acts on it is SL(2,R), whose generators we denote by
Pauli matrices  $\tau_1, \tau_3$.  (We will mostly avoid using $i\tau_2$, which
corresponds to the compact generator.) 

\subsection{Global Super-Poincar\'e Symmetry}

World-volume coordinates are denoted $\sigma^\mu, $ $\mu = 0, 1, \ldots, p$. 
The world-volume signature is also taken to be $(-+ \cdots +)$. The
world-volume theory is supposed to have global IIA or IIB super-Poincar\'e
symmetry.  This is achieved by constructing it out of manifestly supersymmetric
quantities.  The superspace supersymmetry transformations are
\begin{equation}
\delta_\epsilon \theta = \epsilon, \qquad \delta_\epsilon X^m = \bar\epsilon \Gamma^m
\theta. \label{susytrans}
\end{equation}
Thus it is evident that two supersymmetric quantities
are $\partial_\mu \theta$  and
\begin{equation}
\Pi_\mu^m = \partial_\mu X^m - \bar\theta \Gamma^m \partial_\mu \theta.
\end{equation}
Another useful quantity is the induced world-volume metric
\begin{equation}
G_{\mu\nu} = \eta_{mn} \Pi_\mu^m \Pi_\nu^n.
\end{equation}
This is also supersymmetric, of course.

Many subsequent formulas are written more concisely
using differential form notation. In doing this one has to be careful about minus signs
when Grassmann variables appear. The basic rule that we use is that
$d = d\sigma^{\mu} \partial_{\mu}$ and $d\sigma^{\mu}$ is regarded
as an odd element of the Grassmann algebra. Thus, for example,
\begin{equation}
d \theta = d\sigma^{\mu} \partial_{\mu} \theta = -  \partial_{\mu} \theta d\sigma^{\mu}.
\end{equation}
There are various possible conventions that would be consistent. Ours, while
perhaps not the most common, is convenient. Taking $\theta$ to anticommute
with $d\sigma^{\mu}$ allows us to keep track of just one (combined)
grading instead of two.
In this notation, two supersymmetric one-forms are  $d\theta$
and $\Pi^m = dX^m  + \bar\theta \Gamma^m d\theta$. We also have
\begin{equation}
d \Pi^m = d \bar\theta \Gamma^m d \theta.
\end{equation}
Wedge products are always implicit in our formulas.

D-brane theories also contain a world-volume gauge field, and so we require
a suitable supersymmetric expression in which it appears.
This turns out to be
\begin{equation}
{\cal F}_{\mu\nu} = F_{\mu\nu} - b_{\mu\nu},
\end{equation}
where $F_{\mu\nu} = \partial_\mu A_\nu - \partial_\nu A_\mu$.  
The structure of ${\cal F}_{\mu\nu}$ is most easily described in terms of the
2-form ${\cal F} = {1 \over 2}{\cal F}_{\mu\nu} d \sigma^\mu
d\sigma^\nu$.  Then ${\cal F} = F - b$, and the appropriate choice of $b$ turns
out to be~\cite{townsend}
\begin{equation}
b = - \bar\theta \Gamma_{11} \Gamma_m d\theta \left(d X^m + {1\over 2} \bar
\theta \Gamma^m d\theta\right). \label{bformula}
\end{equation}
In components the formula for $b$ becomes
\begin{equation}
b_{\mu\nu} = \bar\theta \Gamma_{11} \Gamma_m \partial_{\mu}\theta 
\left(\partial_{\nu} X^m - {1\over 2} \bar
\theta \Gamma^m \partial_{\nu}\theta\right) - (\mu \leftrightarrow \nu).
\end{equation}
This is the formula for $p$ even.  When $p$ is odd, $\Gamma_{11}$ is replaced by
$\tau_3$.  A crucial feature of this choice of $b$ is that $\delta_\epsilon b$ is an
exact differential form. 
This implies that ${\cal F}$ is supersymmetric for an appropriate choice
of $\delta_{\epsilon} A$. 

To be explicit, using eq. (\ref{susytrans})
\begin{equation}
\delta_\epsilon b = -\bar\epsilon \Gamma_{11} \Gamma_m d\theta \left(d X^m +
{1\over 2} \bar\theta \Gamma^m d\theta\right)
+ {1 \over 2} \bar\theta \Gamma_{11} \Gamma_m d\theta \bar\epsilon \Gamma^m
d\theta.
\end{equation}
To show that this is an exact 
differential form, we substitute $\theta = \theta_1 + \theta_2$, using 
eq. (\ref{thetas}). This gives
\begin{equation}
\delta_{\epsilon} b = (\bar\epsilon_1 \Gamma_m d \theta_1 - 
\bar\epsilon_2\Gamma_m d\theta_2) dX^m +
\bar\epsilon_1\Gamma_m d\theta_1 \bar\theta_1 \Gamma^m d\theta_1
- \bar\epsilon_2\Gamma_m d\theta_2 \bar\theta_2 \Gamma^m d\theta_2.
\end{equation}
The next step is to use the following fundamental identity, 
which is valid for any three Majorana--Weyl
spinors $\lambda_1, \lambda_2, \lambda_3$ of the same chirality, 
\begin{equation}
\Gamma_m \lambda_1 \bar\lambda_2 \Gamma^m \lambda_3 + \Gamma_m \lambda_2
\bar\lambda_3 \Gamma^m \lambda_1 + \Gamma_m \lambda_3 \bar\lambda_1 \Gamma^m
\lambda_2 = 0. \label{3spinors}
\end{equation}
This formula is valid regardless of whether each of the $\lambda$'s is an even element
or an odd element of the Grassmann algebra.  (Note that $\theta$ is odd and
$d\theta$ is even.)  It implies, in particular, that
\begin{equation}
\bar\epsilon_1\Gamma_m d\theta_1 \bar\theta_1 \Gamma^m d\theta_1
= -{1\over 2} \bar\epsilon_1\Gamma_m \theta_1 d\bar\theta_1 \Gamma^m d\theta_1
=  -{1 \over 3} d\Big(\bar\epsilon_1\Gamma_m \theta_1 \bar\theta_1 \Gamma^m d\theta_1\Big).
\end{equation}
Thus $\delta_\epsilon {\cal F} = 0$ if we take
\begin{equation}
\delta_\epsilon A =  \bar\epsilon \Gamma_{11} \Gamma_m \theta d X^m + {1\over
6} (\bar\epsilon \Gamma_{11} \Gamma_m \theta \bar\theta \Gamma^m d\theta +
\bar\epsilon \Gamma_m \theta\bar\theta \Gamma_{11} \Gamma^m  d\theta).
\label{Asusy1}
\end{equation}

\medskip

\section{Gauge-Invariant D-Brane Actions}

As in the case of super $p$-branes, the world-volume theory 
of a D-brane is given by a sum of
two terms $S = S_1 + S_2$. The first term
\begin{equation}
S_1 = - \int d^{p + 1} \sigma \sqrt{- {\rm {\rm det}} (G_{\mu\nu} + {\cal F}_{\mu\nu})}
\end{equation}
is essentially an amalgam of the Born--Infeld and Nambu--Goto formulas. 
Since it is constructed entirely out of $\Pi^m$ and ${\cal F}$, it is
manifestly supersymmetric. The second term
\begin{equation}
S_2 = \int \Omega_{p + 1},
\end{equation}
where $\Omega_{p + 1}$ is a $(p + 1)$-form, is a Wess--Zumino-type term.  
It describes the coupling of the Ramond-Ramond background field strengths to the
D-brane. It is a characteristic feature of D-branes, of course, that they carry an RR
charge~\cite{polchinski1}.

To understand the supersymmetry of $S_2$, it is useful 
to characterize it by a formal $(p+2)$-form
\begin{equation} 
I_{p+2} = d\Omega_{p+1}, 
\end{equation}
which is a very standard thing to do for Wess-Zumino terms. The advantage of this
is that $I_{p+2}$ is typically a much simpler and more symmetrical expression
than $\Omega_{p+1}$. In particular, we will eventually show that it can be
constructed entirely from the supersymmetry invariants $d\theta$, $\Pi^m$,
and ${\cal F}$. This implies that the variation  $\delta_{\epsilon} \Omega_{p+1}$
must be exact, and therefore $S_2$ is invariant. (We are assuming that there is no 
non-trivial cohomology.)

The two terms $S_1$ and $S_2$ are also
manifestly invariant under $(p + 1)$-dimensional general coordinate
transformations.  However, the other crucial local symmetry -- local kappa 
symmetry  -- will be achieved by a subtle
conspiracy between them, just as in the case of super $p$-branes.

\subsection{Local Kappa Symmetry}

Under local kappa symmetry the variation $\delta\theta$ will be restricted in a
way that will be determined in the course of the analysis.  However, we require that,
whatever $\delta\theta$ is,
\begin{equation}
\delta X^m = \bar\theta \Gamma^m \delta\theta = - \delta \bar\theta \Gamma^m \theta,
\label{Xtrans}
\end{equation}
just as for super $p$-branes.  It follows that
\begin{equation}
\delta\Pi_\mu^m = - 2\delta \bar\theta \Gamma^m \partial_\mu \theta.
\end{equation}
Equivalently, one can write
\begin{equation}
\delta\Pi^m = 2\delta \bar\theta \Gamma^m d \theta.
\end{equation}
Another useful definition is the ``induced $\gamma$ matrix''
\begin{equation}
\gamma_\mu \equiv \Pi_\mu^m \Gamma_m.
\end{equation}
Note that $\{\gamma_\mu, \gamma_\nu\} = 2 G_{\mu\nu}$.  These formulas imply that
\begin{equation}
\delta G_{\mu\nu} = - 2\delta \bar\theta (\gamma_\mu \partial_\nu + \gamma_\nu
\partial_\mu)\theta. \label{Gtrans}
\end{equation}

To derive the formula
for  $\delta {\cal F}$, one first uses eqs. (\ref{bformula}) and (\ref{Xtrans}) to compute
\begin{equation}
\delta b =  -2 \delta \bar\theta \Gamma_{11} \Gamma_m d\theta \Pi^m
+ d\Big(-\delta\bar\theta \Gamma_{11} \Gamma_m \theta \Pi^m + {1\over 2}
\delta \bar\theta \Gamma_{11} \Gamma_{m} \theta \bar \theta \Gamma^m d\theta
- {1\over 2} \delta \bar\theta \Gamma^m \theta \bar\theta \Gamma_{11}
\Gamma_m d\theta\Big).
\end{equation}
Then, to obtain a relatively simple result for $\delta {\cal F}$,  let us require that
\begin{equation}
\delta A = - \delta\bar\theta \Gamma_{11} \Gamma_m \theta \Pi^m + {1\over 2}
\delta \bar\theta \Gamma_{11} \Gamma_{m} \theta \bar \theta \Gamma^m d\theta
- {1\over 2} \delta \bar\theta \Gamma^m \theta \bar\theta \Gamma_{11}
\Gamma_m d\theta. \label{Atrans}
\end{equation}
The variation of ${\cal F}$ under a kappa transformation
is then
\begin{equation}
\delta {\cal F} = 2 \delta \bar\theta \Gamma_{11} \Gamma_m d\theta \Pi^m,
\end{equation}
or in terms of  components
\begin{equation}
\delta {\cal F}_{\mu\nu} = 2 \delta \bar\theta \Gamma_{11}(\gamma_\mu \partial_\nu -
\gamma_\nu \partial_\mu)\theta. \label{Ftrans}
\end{equation}
These are the formulas for the Type IIA case
($p$ even). In the Type IIB case ($p$ odd), 
one should make the replacement $\Gamma_{11} \rightarrow \tau_3$. 
($\Gamma_{11}$ must not be anticommuted past another $\Gamma$ matrix
before making this substitution!) The
normalization of the two-form $b$ in the preceding section was chosen
so that the formula for $\delta {\cal F}$ obtained in this way would combine nicely
with the formula for $\delta G$ in eq. (\ref{Gtrans}). 
\medskip

\subsection{Determination of $S_2$}

Now let's  consider a kappa transformation of $S_1$ using 
\begin{equation}
\delta L_1=\delta \left(-\sqrt{-{\rm det} (G + {\cal F})}\right) 
= - {1\over 2} \sqrt{-{\rm det} (G +
{\cal F})} {\rm tr} [(G + {\cal F})^{-1} (\delta G + \delta {\cal F})]
\end{equation}
Inserting the variations
$\delta G_{\mu\nu}$ and $\delta {\cal F}_{\mu\nu}$ given in eqs. (\ref{Gtrans}) and
(\ref{Ftrans}) gives
\begin{equation}
\delta L_1
= 2 \sqrt{- {\rm det} (G + {\cal F})} \delta \bar\theta \gamma_\mu \{(G - {\cal F}
\Gamma_{11})^{-1}\}^{\mu\nu} \partial_\nu \theta.
\end{equation}
For $p$ odd $\Gamma_{11}$ is replaced this time by $-\tau_3$, since it has
been moved past $\gamma_{\mu}$.  Now the key step is to rewrite
this in the form
\begin{equation}
\delta L_1 = 2\delta \bar\theta \gamma^{(p)} T_{(p)}^\nu \partial_\nu \theta,
\label{deltaL1}
\end{equation}
where
\begin{equation}
\big(\gamma^{(p)}\big)^2 = 1. \label{gammasquare}
\end{equation}
It is not at all obvious that this is possible.  The proof that it is, and the
simultaneous determination of $\gamma^{(p)}$ and $T_{(p)}^\nu$, will
emerge as we proceed.

It is useful to define
\begin{equation}
\rho^{(p)} = \sqrt{- {\rm det} (G + {\cal F})} \gamma^{(p)},
\end{equation}
so that eq. (\ref{gammasquare}) becomes
\begin{equation}\label{1star}
\big(\rho^{(p)}\big)^2 = - {\rm det} (G + {\cal F}).
\end{equation}
The requirement
\begin{equation}
\sqrt{-{\rm det} (G+{\cal F})} \gamma_\mu \{(G - {\cal F}
\Gamma_{11})^{-1}\}^{\mu\nu} = \gamma^{(p)} T_{(p)}^\nu
\label{referinD}
\end{equation}
can then be recast in the more convenient form
\begin{equation}\label{2star}
\rho^{(p)} \gamma_\mu = T_{(p)}^\nu (G - {\cal F} \Gamma_{11})_{\nu\mu}.
\end{equation}

It is also useful to represent $\rho^{(p)}$ in terms of an antisymmetric tensor
\begin{equation}
\rho^{(p)} = {1\over(p + 1)!}{\epsilon^{\mu_{1}\mu_{2}\ldots \mu_{p+1}}}
\rho_{\mu_{1}\mu_{2}\ldots \mu_{p+1}}, \label{rhoup}
\end{equation}
or, equivalently, by a $(p + 1)$-form
\begin{equation}
\rho_{p+1} = {1\over (p + 1)!}{\rho_{\mu_{1}\mu_{2}\ldots \mu_{p+1}}}
d\sigma^{\mu_{1}} d\sigma^{\mu_{2}}\ldots d\sigma^{\mu_{p+1}}.
\label{rhodown}
\end{equation}
Similarly, the vector $T_{(p)}^\nu$ can be represented by an antisymmetric tensor
\begin{equation}
T_{(p)}^{\nu} = {1\over p!}{\epsilon^{\nu_{1}\nu_{2}\ldots\nu_{p} \nu}}
T_{\nu_{1}\nu_{2}\ldots\nu_{p}},
\label{Tup}
\end{equation}
or, equivalently, by a $p$-form
\begin{equation}
T_p = {1\over p!}{T_{\nu_{1}\nu_{2}\ldots\nu_{p}}} d\sigma^{\nu_{1}}
\ldots d\sigma^{\nu_{p}}.
\label{Tdown}
\end{equation}

In order to achieve kappa symmetry, we require that
\begin{equation}
\delta L_2 = 2\delta \bar\theta T^\nu_{(p)} \partial_\nu \theta,
\end{equation}
so that adding eq. (\ref{deltaL1}) gives
\begin{equation}
\delta (L_1 + L_2) = 2\delta \bar\theta (1 + \gamma^{(p)}) T^\nu_{(p)} \partial_\nu
\theta. \label{totalvar}
\end{equation}
Since ${1\over 2} (1 \pm \gamma^{(p)} )$ are projection operators,
$\delta \bar\theta = \bar \kappa (1 - \gamma^{(p)}) $ gives
the desired symmetry.
In terms of differential forms, the kappa variation of $S_2$ is
\begin{equation}
\delta S_2 = 2 (-1)^{p+1}\int \delta \bar\theta T_p d\theta = \delta \int \Omega_{p+1}.
\end{equation}
The preceding formula and the definition $I_{p+2} = d\Omega_{p+1}$ implies that
\begin{equation}
\delta I_{p+2} = 2(-1)^{p+1}d\Big( \delta\bar\theta T_p d\theta\Big) = 
2 (-1)^{p+1}\Big( \delta d\bar\theta T_p d\theta -  \delta\bar\theta dT_p d\theta\Big).
\end{equation}
This equation is solved by
\begin{equation}
I_{p+2} = (-1)^{p+1}d\bar\theta T_p d\theta,
\label{WZformula}
\end{equation}
since we will show that
\begin{equation}\label{3star}
d\bar\theta\delta T_p d\theta + 2 \delta \bar\theta d T_p d\theta = 0.
\end{equation}
A corollary of
this identity is the closure condition $d I_{p+2} = d \bar\theta dT_p d\theta =0.$

\subsection{Solution of  Eqs.~(\ref{1star}) and~(\ref{2star})}

Let us now present the solution of eqs.~(\ref{1star}) and~(\ref{2star}) first
for the case of $p$ even.  For this purpose we define the matrix-valued
one-form
\begin{equation}
\psi \equiv \gamma_\mu d\sigma^\mu = \Pi^m \Gamma_m,
\end{equation}
and introduce the following formal sums of differential forms (the subscript $A$ denotes IIA)
\begin{equation}
\rho_A = \sum_{p={\rm even}} \rho_{p+1}
 \quad {\rm and} \quad T_A = \sum_{p={\rm even}} T_p. \label{rhoandT}
\end{equation}
Then the solution of eqs.~(\ref{1star}) and~(\ref{2star}) is described by the
formulas
\begin{equation}
\rho_A = e^{{\cal F}} S_A (\psi) \quad {\rm and} \quad
T_A = e^{{\cal F}} C_A (\psi),
\end{equation}
where
\begin{equation}
S_A (\psi) = \Gamma_{11} \psi + {1\over 3!} \psi^3 + {1\over 5!} \Gamma_{11}
\psi^5 + {1\over 7!} \psi^7 + \ldots \label{SAeqn}
\end{equation}
\begin{equation}
C_A (\psi) = \Gamma_{11} + {1\over 2!} \psi^2 + {1\over 4!} \Gamma_{11} \psi^4
+ {1\over 6!} \psi^6 + \ldots \quad . \label{CAeqn}
\end{equation}
Thus, 
\begin{equation}
\rho_1 = \Gamma_{11} \psi, \quad \rho_3 = {1\over 6} \psi^3 + {\cal F}
\Gamma_{11} \psi, \quad \rho_5 =  {1\over 120} \Gamma_{11} \psi^5
+  {1 \over 6} {\cal F} \psi^3 + {1 \over 2} {\cal F}^2 \Gamma_{11} \psi, \quad {\rm etc.},
\end{equation}
and 
\begin{equation}
T_0 = \Gamma_{11}, \quad T_2 = {1\over 2} \psi^2 + {\cal F} \Gamma_{11}, 
\quad T_4 = {1 \over 24} \Gamma_{11} \psi^4 + {1 \over 2} {\cal F} \psi^2  
+ {1 \over 2} {\cal F}^2 \Gamma_{11}, \quad {\rm etc.}
\end{equation}
Note that the Wess-Zumino term $S_2 $ is given by
$I_2 = -d\bar\theta \Gamma_{11} d \theta$ for the D 0-brane.  The
fact that it is nonvanishing, in contrast to the superparticle of ~\cite{brink}, can be traced
to the fact that the
D 0-brane is massive, whereas the superparticle considered in~\cite{brink} is massless.
The proofs that these expressions for $\rho_A$ and $T_A$ satisfy 
eqs.~(\ref{1star}),~(\ref{2star}) and~(\ref{3star}) are given in
Appendices A,B, and C, respectively. 

Separating positive chirality ($\theta_1$) and negative chirality ($\theta_2$) subspaces,
$\rho_A$ and $T_A$ can be rewritten as $2 \times 2$ matrices
\begin{equation}
\rho_A = e^{\cal F} \left(\begin{array}{cc}
0 & {\rm sinh}\, \psi\\
- {\rm sin}\, \psi & 0\end{array}\right)
\label{rhoA2by2}
\end{equation}
and
\begin{equation}
T_A = e^{\cal F} \left(\begin{array}{cc}
{\rm cosh}\, \psi & 0\\
0 & - {\rm cos} \, \psi \end{array}\right).
\label{TA2by2}
\end{equation}

The solution for $p$ odd is very similar.  In this case we define (the subscript $B$ denotes IIB)
\begin{equation}
\rho_B = \sum_{p={\rm odd}} \rho_{p+1}\quad {\rm and} \quad
T_B = \sum_{p={\rm odd}} T_p. \label{rhoB}
\end{equation}
The solution is given by
\begin{equation}
\rho_B = e^{{\cal F}} C_B (\psi)\tau_1 \quad {\rm and} \quad
T_B = e^{{\cal F}} S_B (\psi)\tau_1 ,
\end{equation}
where
\begin{equation}
S_B (\psi) = \psi + {1\over 3!} \tau_3 \psi^3 + {1\over 5!} \psi^5 + {1\over
7!} \tau_3 \psi^7 +\ldots 
\end{equation}
\begin{equation}
C_B (\psi) = \tau_3 + {1\over 2!}\psi^2 + {1\over 4!} \tau_3 \psi^4 + {1\over
6!} \psi^6 +\ldots \quad. \label{CBeqn}
\end{equation}
Thus 
\begin{equation}
\rho_2 = {1\over 2} \tau_1 \psi^2 + i \tau_2 {\cal F}, \quad \rho_4 = 
{1 \over 24} i \tau_2
\psi^4 + {1 \over 2} \tau_1 {\cal F} \psi^2 + {1 \over 2} i \tau_2 {\cal F}^2, \quad {\rm etc.}, 
\end{equation}
and 
\begin{equation}
T_1
= \tau_1 \psi, \quad T_3 = {1 \over 6} i \tau_2 \psi^3 + \tau_1 {\cal F} \psi, \quad {\rm etc.}  
\end{equation}
The quantity $\rho_0 = i \tau_2$ may be relevant to the D-instanton, which we are
not considering here.
The proofs that these expressions for $\rho_B$ and $T_B$  satisfy 
eqs.~(\ref{1star}),~(\ref{2star})  (with $\Gamma_{11}$ replaced
by $-\tau_3$) and~(\ref{3star}) are essentially the same as in the IIA case (see
Appendices A, B, and C). 

Displaying the Pauli matrices explicitly,
$\rho_B$ and $T_B$ can be rewritten as $2 \times 2$ matrices
\begin{equation}
\rho_B = e^{\cal F} \left(\begin{array}{cc}
0 & {\rm cosh} \, \psi\\
- {\rm cos} \, \psi & 0\end{array}\right).
\label{rhoB2by2}
\end{equation}
and
\begin{equation}
T_B = e^{\cal F} \left(\begin{array}{cc}
0 & {\rm sinh} \, \psi\\
{\rm sin} \, \psi & 0\end{array}\right).
\label{TB2by2}
\end{equation}

\subsection{A Comment on Conventions}

Because of certain automorphisms, some arbitrary choices had to be made in the
preceding formulas. For example, $S_1$ is invariant under ${\cal F} \to - {\cal F}$.
The associated freedom was resolved by choosing the $\rho$'s and $T$'s to
contain the factor ${\rm exp} ({\cal F})$ rather than  ${\rm exp} (-{\cal F})$.
One also has the freedom to replace $(\rho, T)$ by $(-\rho, -T)$ for any set of $p$'s.
Other arbitrary choices in the IIA case stem from the automorphisms of the
Dirac algebra $\Gamma_m \to - \Gamma_m$ and $\Gamma_{11} \to
- \Gamma_{11}$. This freedom was resolved by choosing all of the coefficients
in the expansions of $S_A$ and $C_A$ to be positive. Similarly, in the IIB case
the automorphisms are $\Gamma_m \to - \Gamma_m$ and the $SO(2)$ group
of automorphisms of $SL(2,R)$ given by rotating $\tau_1$ and $\tau_3$.
This freedom was resolved up to a pair of minus signs by choosing to use $\tau_3$ in
eq. (\ref{Ftrans}). The remaining signs were settled by the choice of coefficients
in $S_B$ and $C_B$.

\subsection{The Algebra of Kappa Transformations}

It is natural to explore the commutator of two kappa symmetries to determine
whether all the local symmetries have been identified, or whether additional
ones are generated.  In the case of the superstring, it was claimed that a ``new''
local bosonic symmetry is generated~\cite{green1}.
Actually, as we will explain in the context of D-branes, this is not the case.

We have computed the commutator of two kappa transformations for the IIA
$p$-brane theories.  The result is that closure of the algebra requires using the
equation of motion
\begin{equation}
(1 + \gamma^{(p)}) T_{(p)}^\nu \partial_\nu \theta = 0,
\end{equation}
which can be inferred from eq. (\ref{totalvar}).  Then the commutator gives the sum of a
general coordinate transformation and a local kappa symmetry transformation.

Let us recall why it is legitimate to drop the equation of motion terms in the
symmetry algebra.\footnote{We are grateful to R. Kallosh for
explaining this to us.} To explain the general idea,
consider a theory with fields $\Phi^i$ and action $S[\Phi]$,
so that the equations of motion are ${\delta S\over \delta \Phi_i} = 0$.  In
this case, a transformation of the form
\begin{equation}
\delta \Phi_i = \Sigma_{ij} {\delta S\over \delta \Phi_j},
\end{equation}
gives a variation
\begin{equation}
\Sigma_{ij} {\delta S\over\delta\Phi_i} {\delta S\over\delta \Phi_j},
\end{equation}
which vanishes by symmetry for $\Sigma_{ij} = - \Sigma_{ji}$.  Clearly, such a
symmetry can have no significant dynamical implications.  This is precisely what must
always happen for the equation of motion terms in a symmetry
algebra, since the commutator of two symmetry transformations is
necessarily a symmetry transformation. This can be verified explicitly for the
equation of motion term described above.

\medskip

\section{The Static Gauge}

In theories with gauge symmetries, such as those described here, it is often
worthwhile to use the local symmetries to make a gauge choice that
eliminates unphysical degrees of freedom.
In general, there are many consistent choices that can be made, but some
choices turn out to be much more convenient than others.  This is certainly the
case for super D-branes. If we want to obtain tractable formulas, it is
essential to make a good choice.

In the cases $p = 0$ and $p = 1$, there exist special gauge choices that reduce
the equations of motion to free field equations.  In the case of $p=0$, one sets
$G_{00} = -1$, together with a suitable fermionic condition 
such as $\Gamma^+\theta = 0$. In the $p=1$ case, one sets $G_{\mu\nu}$
proportional to the Lorentz metric on the world volume $\eta_{\mu\nu}$,
and imposes a fermionic condition like that of the  $p=0$ case.
The problem with these gauge choices is that they do not have a natural
generalization to $p\geq 1$ that preserves world-volume Lorentz
invariance.\footnote{For a discussion of super $p$-branes in the light-cone gauge
see refs.~\cite{dewit,bergshoeff5}.}
Therefore, in order to have a prescription that applies to all
D-branes at once, we will consider a ``static gauge''
choice instead.  In this gauge the world-volume general coordinate invariance
is used to equate $p + 1$ of the target-space coordinates with the world-volume
coordinates $(X^\mu = \sigma^\mu)$.\footnote{This is only possible when there
are no global topological obstructions to this identification.}
The real challenge is to find a fermionic
gauge condition to supplement this in the case of super D-branes. The price
one pays for the universality of the prescription is that it is not apparent that
the $p=0$ and $p=1$ cases are actually free theories.

\subsection{The Bosonic D-Brane}

To illustrate the issues we wish to address in connection with gauge fixing, it
is helpful to consider first the simpler problem of a ``bosonic D-brane.''  By
this we simply mean the theory obtained by dropping the $\theta$ coordinates
from the supersymmetric D-brane actions.  Thus
\begin{equation}
S_B = - \int \sqrt{-\det (G_{\mu\nu} + F_{\mu\nu})} d^{p + 1} \sigma,
\end{equation}
where
\begin{equation}
G_{\mu\nu} = \eta_{mn} \partial_\mu X^m \partial_\nu X^n.
\end{equation}
The equation of motion obtained by varying $X^m$ is
\begin{equation}
\partial_\mu \left(\sqrt{-\det (G+F)} \left\{(G+F)^{-1} +
(G-F)^{-1}\right\}^{\mu\nu} \partial_\nu X^m \right) = 0.
\end{equation}
Similarly, the $A_\nu$ equation of motion is
\begin{equation}
\partial_\mu \left(\sqrt{-\det (G+F)} \left\{(G+F)^{-1} -
(G-F)^{-1}\right\}^{\mu\nu} \right) = 0.
\end{equation}

The general coordinate invariance of $S_B$ allows us to choose a static gauge
in which the first $p + 1$ components of $X^m$ are equated to the world-volume
coordinates $\sigma^\mu$.  The remaining components of $X^m$ will be denoted
$\phi^i$.  In this gauge $G_{\mu\nu}$ becomes
\begin{equation}
G_{\mu\nu} = \eta_{\mu\nu} + \partial_\mu \phi^i \partial_\nu \phi^i.
\end{equation}
The first $p+1$ of the  $X$ equations become
\begin{equation}\label{Xeqn}
\partial_\mu \left(\sqrt{-\det(G+F)} \left\{(G+F)^{-1} + (G-F)^{-1}
\right\}^{\mu\nu} \right) = 0,
\label{missing}
\end{equation}
and the remaining $\phi^i$ equations become (as a result)
\begin{equation}
\{(G+F)^{-1}\}^{\mu\nu} \partial_\mu \partial_\nu \phi^i = 0.
\end{equation}
Now consider the action obtained by substituting the gauge conditions directly
into $S_B$.  This produces
\begin{equation}
S'_B = - \int \sqrt{-\det (\eta_{\mu\nu} + \partial_\mu \phi^i \partial_\nu
\phi^i + F_{\mu\nu})} d^{p+1} \sigma.
\end{equation}
Varying the $\phi^i$ and $A_\mu$ variables in this action certainly gives the
correct gauge-fixed form of their field equations.  The more subtle question is
whether the additional $p + 1$ equations in eq. (\ref{Xeqn}),
which are associated to degrees of
freedom that do not appear in $S'_B$, need to be imposed as additional
constraints or  whether they are automatically satisfied.

To decide on the status of eq. (\ref{Xeqn}), it is helpful to observe that it has the
form $\partial_\mu \Theta^{\mu\nu} = 0$, where $\Theta^{\mu\nu}$ is the
energy-momentum tensor one would form by replacing $\eta_{\mu\nu}$ by $g_{\mu\nu}$
in $S'_B$, varying with respect to $g_{\mu\nu}$, and then setting it equal to
$\eta_{\mu\nu}$.  It is related to the energy-momentum tensor $T^{\mu\nu}$
that one would form as the Noether current for the world-volume
translational symmetries by an equation of the form $\Theta^{\mu\nu} = T^{\mu\nu}
+ \partial_{\lambda} \Sigma^{\mu\nu\lambda}$, where 
$\Sigma^{\mu\nu\lambda} = -\Sigma^{\lambda\nu\mu}$.
This is conserved, of course, as a consequence of the
equations of motion of $\phi^i$ and $A_\mu$.  
Thus $S'_B$ encodes all of the dynamics, and no
supplementary constraints are required.

Among the symmetries of the gauge-invariant action $S_B$ were the global
translations symmetries $\delta X^m = a^m$.  Decomposing these into
translations $a^\mu$ tangent to the brane and $a^i$ orthogonal to the  brane, we have
\begin{equation}
\delta X^\mu = a^\mu + \xi^\rho \partial_\rho X^\mu = a^\mu + \xi^\mu = 0
\end{equation}
and
\begin{equation}
\delta \phi^i = a^i + \xi^\rho \partial_\rho \phi^i = a^i - a^\rho
\partial_\rho \phi^i.
\end{equation}
In these equations we have added a compensating general coordinate
transformation, with parameter $\xi^\mu = - a^\mu$, in order to preserve the
$X^\mu = \sigma^\mu$ gauge.  One can also deduce the induced transformation of
the gauge field $A_\mu$.  Up to a gauge transformation, it is
\begin{equation}
\delta A_\mu = - a^\rho F_{\rho\mu}.
\end{equation}

The symmetry $\delta \phi^i = a^i$ is a trivial, but significant, symmetry of
the theory.  It is the translational symmetry orthogonal to the brane, which is
broken as a consequence of the presence of the brane.  The inhomogeneity of
$\delta \phi^i = a^i$ is the signal that we are dealing with a spontaneously
broken symmetry, and that the $\phi^i$ are the associated Goldstone bosons.
The unbroken translation symmetries tangent to the brane have parameters
$a^\rho$.  This symmetry is just the obvious translation symmetry of the gauge
fixed world-volume theory.

\subsection{Super D-Branes in Static Gauge}

In the case of super D-branes we will again use the static gauge choice $X^\mu
= \sigma^\mu$, described in the preceding subsection, and
denote the $9-p$ remaining $X$ coordinates by $\phi^i$. 
The crucial question is how to use the kappa symmetry to eliminate half of the
components of the $\theta$ coordinates. We have discovered 
a natural choice that leads to surprisingly simple and tractable formulas:
we set one of the two Majorana--Weyl spinors that comprise $\theta$ equal to zero.
Specifically, in the IIA case we set $\theta_2
= 0$. The surviving Majorana--Weyl spinor, $\theta_1$,
 is then renamed $\lambda$. We could do the same thing for the IIB case.
However, for the conventions that have been introduced in the preceding
sections, it is more convenient to set $\theta_1 =0$ and $\theta_2 = \lambda$
in the IIB case. This will result in formulas that take the same form in both cases.

After gauge fixing, $\lambda$ becomes a world-volume spinor, 
even though the $\theta$'s
were originally world-volume scalars, as a consequence of the identification
of $X^{\mu}$ with $\sigma^{\mu}$. When $p\leq 5$, the gauge-fixed world-volume 
theory has extended supersymmetry. In this case the 32-component 
Majorana--Weyl spinor $\lambda$ actually represents a set of minimal spinors.
However, we find it convenient to leave it alone rather than to decompose
it into pieces.

To see how our proposed gauge choice works, let us consider the global
supersymmetry transformations with parameter $\epsilon$, which we now decompose
into two parts called $\epsilon_1$ and $\epsilon_2$.  In the IIA case
$\epsilon_1$ and $\epsilon_2$ have opposite chirality, while in the IIB case
they have the same chirality.  Nonetheless, all the formulas that follow are
valid for both cases (unless otherwise indicated).  Since supersymmetry
transformations move the variables out of the gauge, it is necessary to add
compensating general coordinate and kappa transformations that restore the
gauge\footnote{Exactly the same sort of reasoning can be used to argue that
the gauge choice is consistent in the first place.}
\begin{eqnarray}
\delta \bar\theta &=& \bar\epsilon + \bar\kappa (1 - \gamma^{(p)}) + \xi^\mu
\partial_\mu \bar\theta \nonumber \\
\delta X^m &=& \bar\epsilon \Gamma^m \theta - \bar\kappa (1 - \gamma^{(p)})
\Gamma^m \theta + \xi^\mu \partial_\mu X^m.
\end{eqnarray}

{}From the structure of $\rho_A$ and $\rho_B$ displayed in
eqs. (\ref{rhoA2by2}) and (\ref{rhoB2by2}), one sees 
that the matrix $\gamma^{(p)}$ is
off-diagonal in both the IIA and IIB cases.  Thus we write it in the block form
\begin{equation}
\gamma^{(p)} = \left(\begin{array}{cc}
0 &\zeta^{(p)} \\
\tilde{\zeta}^{(p)} & 0\end{array}\right).
\label{gamma2by2}
\end{equation}
The equation $[\gamma^{(p)}]^2 = 1$ then becomes
\begin{equation}
\zeta^{(p)} \tilde{\zeta}^{(p)} = \tilde{\zeta}^{(p)} \zeta^{(p)} = 1.
\end{equation}
There is no reason that the square of $\zeta^{(p)}$ should be
anything simple.  In this notation, the requirements $\delta \bar\theta_2 = 0$ 
and $\delta X^\mu = 0$ in the IIA case become
\begin{eqnarray}
0 &=& \bar\epsilon_2 + \bar\kappa_2 - \bar\kappa_1 \tilde{\zeta}^{(p)}\nonumber
\\
0 &=& \bar\epsilon_1 \Gamma^\mu \lambda - (\bar\kappa_1 - \bar\kappa_2
\zeta^{(p)}) \Gamma^\mu \lambda + \xi^\mu.
\end{eqnarray}
The above equations are solved by
\begin{equation}
\bar\kappa_1 - \bar\kappa_2 \zeta^{(p)} = \bar\epsilon_2 \zeta^{(p)}
\end{equation}
\begin{equation}
\xi^\mu = (\bar\epsilon_2 \zeta^{(p)} - \bar\epsilon_1) \Gamma^\mu \lambda.
\label{xivalue}
\end{equation}
For these choices the total supersymmetry transformations of the fields that
remain in the gauge-fixed theory are
\begin{eqnarray}
\Delta \bar\lambda &=& \bar\epsilon_1 + \bar\epsilon_2 \zeta^{(p)} + \xi^\mu
\partial_\mu \bar\lambda \nonumber \\
\Delta \phi^i &=& (\bar\epsilon_1 - \bar \epsilon_2 \zeta^{(p)}) \Gamma^i
\lambda + \xi^\mu \partial_\mu \phi^i \nonumber \\
\Delta A_\mu &=& (\bar\epsilon_2 \zeta^{(p)} - \bar\epsilon_1) (\Gamma_\mu +
\Gamma_i \partial_\mu \phi^i)\lambda \nonumber \\
& &  + ({1\over 3} \bar\epsilon_1 - \bar\epsilon_2 \zeta^{(p)}) \Gamma_m \lambda
\bar\lambda \Gamma^m \partial_\mu \lambda + \xi^\rho \partial_\rho A_\mu +
\partial_\mu \xi^\rho A_\rho. \label{ptrans}
\end{eqnarray}
Note that the index $m$ is a 10d index, which includes both $\mu$ and $i$
values.  The parameter $\xi^\mu$ in these equations is understood to take the
value given in eq. (\ref{xivalue}).

In the IIB case, one finds exactly the same set of gauge-fixed supersymmetry
transformations except that the labels 1 and 2 on $\epsilon$ and $\kappa$
are interchanged. Since there is no fundamental distinction between them,
anyway, we simply interchange these labels in the IIB case, so that the formulas
then look identical to those of the IIA case.  Thus (\ref{ptrans}) describes
the symmetry transformations of our gauge-fixed theory for all values of $p$
from 0 to 9.

Now let's look at the actions that result from imposing the gauge choices on
the gauge-invariant D-brane actions.  Recall that the Wess--Zumino term
$S_2$ is characterized by the $(p + 2)$-form $I_{p+2} = \pm d \bar\theta T_p
d\theta$.  The crucial fact is that $T_p$ connects $\theta_1 $ and $\theta_2$
in both the IIA and
IIB cases, so that $I_{p+2} \propto d \bar\theta_2 ( \ldots ) d\theta_1$, which
vanishes for $\theta_1=0$ or $\theta_2 = 0$.  
Therefore, only $S_1$ contributes to the gauge
fixed action.  The result is
\begin{equation}
S^{(p)} = - \int d^{p+1} \sigma \sqrt{- {\rm det}\, M^{(p)}},
\label{paction}
\end{equation}
where $M_{\mu\nu}^{(p)} = G_{\mu\nu}^{(p)} + {\cal F}_{\mu\nu}^{(p)}$, and
\begin{equation}
G_{\mu\nu}^{(p)} = \eta_{\mu\nu} + \partial_\mu \phi^i
\partial_\nu \phi^i -  \bar\lambda (\Gamma_\mu + \Gamma_i \partial_\mu
\phi^i)\partial_\nu \lambda
-  \bar\lambda (\Gamma_\nu + \Gamma_i \partial_\nu
\phi^i)\partial_\mu \lambda
+ \bar\lambda \Gamma^m \partial_\mu 
\lambda \bar\lambda \Gamma_m \partial_\nu \lambda,
\end{equation}
\begin{equation}
{\cal F}_{\mu\nu}^{(p)} = F_{\mu\nu} 
-  \bar\lambda (\Gamma_\mu + \Gamma_i \partial_\mu
\phi^i)\partial_\nu \lambda
+ \bar\lambda (\Gamma_\nu + \Gamma_i \partial_\nu
\phi^i)\partial_\mu \lambda.
\end{equation}
Thus
\begin{equation}
M_{\mu\nu}^{(p)} = \eta_{\mu\nu} + F_{\mu\nu} + \partial_\mu \phi^i
\partial_\nu \phi^i - 2 \bar\lambda (\Gamma_\mu + \Gamma_i \partial_\mu
\phi^i)\partial_\nu \lambda
+ \bar\lambda \Gamma^m \partial_\mu 
\lambda \bar\lambda \Gamma_m \partial_\nu \lambda.
\end{equation}

If we have made no errors, the action $S^{(p)}$ 
in eq. (\ref{paction}) should be invariant under the
$\epsilon_1$ and $\epsilon_2$ transformations given  in (\ref{ptrans}).  
However, as a
check of both our reasoning and our mathematics, it is a good idea to check this
explicitly.  In Appendix D we show in detail that the variation of the integrand
of $S^{(p)}$ is a total derivative, as required.  Thus we are confident that
the action and the symmetry transformations are  correct.

The action $S^{(p)}$, which describes a $p$-brane in a flat space-time
background, is (maximally) supersymmetric Born--Infeld theory.  In particular,
for the case $p = 9$, we obtain supersymmetric Born--Infeld theory in 10d.  
In this case there are no transverse coordinates $\phi^i$ and the
target space index $m$ can be replaced by a Greek world-volume index.  The
resulting formula is given by
\begin{equation}
M_{\mu\nu}^{(9)} = \eta_{\mu\nu} + F_{\mu\nu} - 2 \bar\lambda \Gamma_{\mu}
\partial_\nu \lambda + \bar \lambda \Gamma^\rho \partial_\mu \lambda
\bar\lambda \Gamma_\rho \partial_\nu \lambda.
\end{equation}
The supersymmetries of $S^{(9)}$ are given by
\begin{equation}
\Delta \bar\lambda = \bar\epsilon_1 + \bar\epsilon_2 \zeta^{(9)} + \xi^\mu
\partial_\mu \bar\lambda  
\label{lambdasusy}
\end{equation}
\begin{equation}
 \Delta A_\mu = (\bar\epsilon_2 \zeta^{(9)} - \bar\epsilon_1) \Gamma_\mu
\lambda + ({1\over 3} \bar\epsilon_1 - \bar\epsilon_2 \zeta^{(9)}) \Gamma_{\rho}
\lambda \bar\lambda \Gamma^\rho \partial_\mu \lambda
+\xi^\rho \partial_\rho A_\mu + \partial_\mu \xi^\rho A_\rho,
\label{Asusy}
\end{equation}
where
\begin{equation}
\xi^\mu = (\bar\epsilon_2 \zeta^{(9)} - \bar\epsilon_1) \Gamma^\mu \lambda.
\label{inducedxi}
\end{equation}

The transformation with parameter $\epsilon_1$ describes the supersymmetries of
Volkov--Akulov type, which are broken by the presence of the D-brane.   The
inhomogeneity of the $\epsilon_1$ transformation of $\lambda$ shows that it is
the associated Goldstone fermion.  The unbroken supersymmetries should give no
inhomogeneous terms, so they must be given by a combined transformation 
with $\epsilon_1 = \epsilon_2$.  

The ``missing'' equations, analogous to those in eq. (\ref{missing}), 
arise in the IIA case from the gauge-fixed form 
of the $\theta_2$ equation of motion.  All terms in $S_1$ involve
even powers of $\theta_2$, so their $\theta_2$ variations vanish for $\theta_2 =
0$.  However, as we have said, the Wess--Zumino term is characterized by
$I_{p+2} \sim d \bar\theta_2 \big(T_p\big)_{11} d\theta_1$.  
Thus, it contains terms linear in
$\theta_2$.  Indeed, using eq. (\ref{TA2by2}),
the $\theta_2$ equation of motion, evaluated at $\theta_2 = 0$ becomes 
\begin{equation}
\big( e^{\cal F} {\rm cosh}\, \psi\big)_p  d\lambda = 0,
\end{equation}
where it is understood that the $p$-form $\big( e^{\cal F} {\rm cosh}\, \psi\big)_p$ 
is evaluated for $\theta_2 =0$ and $X^{\mu} = \sigma^{\mu}$.  
We know that $I_{p+2}$ is exact, a property that must hold
order-by-order in $\theta_2$.  Using this fact for the linear term, one infers
that $\big( e^{\cal F} {\rm cosh}\, \psi\big)_p d\lambda$ is exact.  Thus we can write
\begin{equation}
\big( e^{\cal F} {\rm cosh}\, \psi\big)_p d\lambda = dJ_p.
\end{equation}
Then the $\theta_2$ equation of motion $dJ_p = 0$ is interpreted as
conservation of the supercurrent $J_p$.  Note that we have
represented the supercurrent $J^\mu_{(p)}$ by a $p$-form $J_p$.  
They are related in the
same way that $T_{(p)}^\mu$ and $T_p$ are in sect. 3.2.  In this notation, the
conserved supercharge is
\begin{equation}
Q^{(p)} = \int_{\Sigma_{p}} J_p,
\end{equation}
where $\Sigma_p$ is a spatial slice of the $(p + 1)$-dimensional world-volume.
The conservation equation $dJ_p = 0$ is a consequence of the equations of
motion of $S^{(p)}$, just as we saw for energy-momentum tensor in the bosonic
theory. The analysis works the same way in the IIB case except that the equation
of motion is
\begin{equation}
\big( e^{\cal F} {\rm sinh}\, \psi\big)_p  d\lambda = dJ_p =0.
\end{equation}
The two cases can be described together by the single equation
\begin{equation}
\big( e^{\cal F+\psi}\big)_p  d\lambda = dJ_p =0.
\end{equation}

\subsection{Dimensional Reduction}

The set of gauge-fixed D-brane actions $S^{(p)}$ given in 
eq. (\ref{paction}) are related to
one another by straightforward dimensional reduction.  In particular, this
means that starting with $S^{(9)}$ and dropping the dependence on $9-p$ of the
world-volume coordinates gives the action $S^{(p)}$.  With our
conventions one must identify the $9 - p$ scalar components of $A$ as $A_i = -
\phi^i$.  (For other conventions this equation could have a plus sign.)

To demonstrate the claim given above, let us consider the dimensional reduction
from $S^{(p)}$ to $S^{(p - 1)}$, so that the general case is implied by
induction.  Setting all $\sigma^p$ derivatives to zero and $A_p = -
\phi^p$, we can write $M^{(p)}$ in block form as
\begin{equation}
M_{\hat\mu \hat\nu}^{(p)} = \left(\begin{array}{l|c}
\eta_{\mu\nu} + F_{\mu\nu} + \partial_\mu \phi^i \partial_\nu \phi^i & \\
- 2 \bar\lambda (\Gamma_\mu + \Gamma_i \partial_\mu \phi^i) \partial_\nu
\lambda & - \partial_\mu \phi^p\\
+ \bar\lambda \Gamma^m \partial_\mu \lambda \bar\lambda \Gamma_m \partial_\nu
\lambda &\\  \\
\hline
\\ -2 \bar\lambda \Gamma_p \partial_\nu \lambda + \partial_\nu \phi^p &
1\end{array} \right).
\end{equation}
It is understood here that $\hat\mu = (\mu, p)$ and the last row and column
correspond to $\hat\mu = p$ and $\hat\nu = p$, respectively.  Also, the index
$i$ is summed from $p+1$ to $9$.  All that matters in the action is the
determinant of $- M^{(p)}$, so we may add multiples of the last row to the
other rows.  Doing this with a factor of $\partial_{\mu} \phi^p$,
so as to create zeros in the upper right corner, one
obtains precisely the desired matrix $M_{\mu\nu}^{(p - 1)}$ in the upper left
block.  The sum on $i$ now goes from $p$ to $9$.
This proves the compatibility of the formulas under dimensional
reduction.

The supersymmetry transformation formulas have the same compatibility under
dimensional reduction.  However, this is a little more work to prove, because one needs
to know a formula for the dimensional reduction of $\zeta^{(p)}$. In Appendix E
we prove that upon dimensional reduction from $p$ to $p-1$ (as above)
\begin{equation}
\zeta^{(p)} \to (-1)^p \Gamma_p \zeta^{(p-1)}.
\end{equation}
This implies that the supersymmetry transformation formulas in
(\ref{ptrans}) retain their form upon dimensional reduction for the
identifications $\bar\epsilon_1^{(p)} = \bar\epsilon_1^{(p-1)}$ and
$\bar\epsilon_2^{(p)} = (-1)^p \bar\epsilon_2^{(p-1)}\Gamma_p$.
The unbroken supersymmetry after dimensional reduction is given by a combined
transformation with $\epsilon_1^{(p)} = \pm \Gamma_{p+1} \ldots 
\Gamma_9 \epsilon_2^{(p)}$. (Some care is required to determine the sign in each case.)

\subsection{The Supersymmetry Algebra}

It is interesting to examine the commutator of two supersymmetry
transformations $\Delta$ and $\Delta'$.  Since lower dimensions can be reached
by dimensional reduction, it is sufficient to consider the 10d case, for which
$\Delta$ is given in eqs. (\ref{lambdasusy}) -- (\ref{inducedxi}).  
(The notation is that $\Delta$ involves
parameters $\epsilon_1$ and $\epsilon_2$, and $\Delta'$ involves parameters
$\epsilon'_1$ and $\epsilon'_2$.)

There are two ways by which the algebra can be computed.  One is to simply do
it directly using eqs. (\ref{lambdasusy}) -- (\ref{inducedxi}).  
This calculation requires knowing
$\Delta\zeta^{(9)}$, which is rather complicated.  The second method is to use
our knowledge of the relationship of $\Delta$ to transformations of 
the gauge-invariant theory.  Let us discuss this approach first.

The $\Delta$ transformation consists of the global supersymmetry
transformation $\delta_\epsilon$ plus compensating general coordinate and
kappa transformations required to maintain the gauge, as described in sect.
4.2.  To infer the commutator of two $\Delta$ transformations, let us first
consider the commutator of two $\delta_\epsilon$ transformations in the 10d ($p
= 9$) gauge-invariant theory.  Using eqs. (\ref{susytrans}) and (\ref{Asusy1}) one obtains
\begin{equation}
[\delta_\epsilon, \delta_{\epsilon'}] \theta = 0\end{equation}
\begin{equation}
[\delta_\epsilon, \delta_{\epsilon'}] X^\mu  = - (a_1^\mu + a_2^\mu)\end{equation}
\begin{equation}
[\delta_\epsilon, \delta_{\epsilon'}] A_\mu  = (a_{2\rho} - a_{1\rho}) \partial_\mu
X^\rho + \partial_\mu \Lambda ,
\end{equation}
where
\begin{equation}
a_1^\mu = 2 \bar\epsilon_1 \Gamma^\mu \epsilon'_1 \quad {\rm and} \quad
a_2^\mu = 2 \bar\epsilon_2 \Gamma^\mu \epsilon'_2
\end{equation}
and
\begin{equation}
\Lambda = \bar\epsilon'_1 \Gamma^\mu \theta_1 \bar\epsilon_2 \Gamma_\mu
\theta_2 - \bar\epsilon_1 \Gamma^\mu \theta_1 \bar\epsilon'_2 \Gamma_\mu
\theta_2.
\end{equation}

Now we claim that the commutator $[\Delta, \Delta']$ in the
gauge-fixed theory can be inferred by taking the $[\delta_\epsilon,
\delta_\epsilon']$ results given above, adding a compensating general
coordinate transformation with parameter $\xi^\mu = a_1^\mu + a_2^\mu$, so that
$X^\mu$ does not transform in the static gauge, and restricting to the static
gauge.
This procedure gives the results (recalling our conventions about $\epsilon_1
\leftrightarrow \epsilon_2, \theta_2 = \lambda$ in gauge-fixing IIB theories)
\begin{equation}
[\Delta, \Delta'] \lambda = (a_1^\mu + a_2^\mu) \partial_\mu \lambda 
\end{equation}
\begin{equation}
[\Delta, \Delta'] A_\mu = a_{1\mu} - a_{2\mu} + 
(a_1^\rho + a_2^\rho) \partial_\rho
A_\mu. \label{Acommutator}
\end{equation}
Thus, we obtain the expected translations together with a constant shift of the
gauge field.  This shift can be regarded as an irrelevant gauge transformation.
 This is true, but on dimensional reduction, it gives rise to shifts of the
scalar fields --- shifts that correspond to the broken translational
symmetries.  Such a shift should not occur in the commutator of two unbroken
supersymmetries.  In fact, $a_{1\mu} - a_{2\mu}$ vanishes for $\epsilon_1 =
\epsilon_2$ and $\epsilon'_1 = \epsilon'_2$, which are the conditions that we
showed earlier describe the unbroken supersymmetries.

The argument we have used to obtain the preceding results depended on the
equivalence of adding compensating gauge transformations before commuting
$\epsilon$ transformations with adding them afterwards.  This procedure must be ambiguous
at least by the type of terms that we have not kept track of in the closure of the
kappa symmetry algebra (see sect. 3.5). This includes terms that vanish on shell,
as well as $U(1)$ gauge transformations. We certainly know that the closure of the 
algebra of unbroken supersymmetries must require an equation of motion, since
$\lambda$ and $A_{\mu}$ do not constitute a complete off-shell multiplet.
With these provisos, we are quite sure that the procedure used to obtain the
commutators given above is correct. However, since it is rather subtle,
we have also carried out some
checks of the algebra $[\Delta, \Delta']$ directly.  The commutator of two
$\epsilon_1$ transformations is quite straightforward and gives the result
stated above without any equation of motion or extra $U(1)$ gauge
transformation.  The commutator of an $\epsilon_1$ and an $\epsilon_2$
transformation requires the identity
\begin{equation}
\Delta (\epsilon_1) \zeta^{(9)} = - \bar\epsilon_1 \Gamma^\mu \lambda
\partial_\mu \zeta^{(9)}.
\end{equation}
Using this, it is straightforward to verify that $[\Delta (\epsilon_1), \Delta
(\epsilon_2) ] = 0$, up to a $U(1)$ gauge transformation,
as required.  The gauge transformation that appears does not contribute
to the transformation of scalar fields after dimensional reduction. So the 
interpretation of the
constant terms that appear in eq. (\ref{Acommutator}) remains as explained
above. The most difficult case is $[\Delta
(\epsilon_2), \Delta (\epsilon'_2)]$, which requires knowing $\Delta (\epsilon_2)
\zeta^{(9)}$.  This is quite complicated, and we have not checked it.

\medskip

\section{Conclusion}

We have presented actions for all type II D-branes $(p = 0, 1, \ldots, 9)$ with
local kappa symmetry in a flat background.  We then fixed a physical gauge by
identifying $p + 1$ of the space-time coordinates with the world-volume
coordinates $(X^\mu = \sigma^\mu)$ and setting one of the two Majorana--Weyl
spinors ($\theta_1$ or $\theta_2$) equal to zero.  This resulted in surprisingly
simple expressions for the gauge-fixed actions.  All of the ones with $p <
9$ can be deduced by dimensional reduction
from the $p = 9$ $(d = 10)$ action given in eq. (1).  This
action thus serves as a master formula for all D-branes.  It is also
interesting as a supersymmetric extension of $d = 10$ Born--Infeld theory.

The results presented here should have a number of applications and
generalizations.  Some of the ones that come to mind are the following: \,1)
studies of $p$-brane dualities in compactified space-times;  \,2) formulation of
non-Abelian generalizations appropriate to the description of systems of
multiple parallel or intersecting D-branes;  \,3) studies of solitons that live
within the D-brane world-volumes;  \,4) formulation of an action for the
M-theory five-brane;  \,5) applications to the study of black holes.

We wish to acknowledge discussions with S. Cherkis, J. Hoppe, R. Kallosh,
D. Lowe, B. Nilsson, and M. Perry.

\newpage
\section*{Appendix A: The Proof of  Equation (\ref{1star})}

We wish to prove that the expressions we have found for $\rho^{(p)}$ satisfy
\[(\rho^{(p)})^{2}=-{\rm det}(G+{\cal F}).\]
The proof is somewhat simpler for the special case  $G_{\mu\nu} = \eta_{\mu\nu}$, 
where $\eta$ is the
flat Minkowski metric with signature $(- + \cdots +)$ in $p+1$ dimensions.
General covariance considerations imply that if 
the formula is true in this case, then it is true in general.
This can be proved, for example, by introducing a vielbein to relate base space
and tangent space coordinates. In the tangent space coordinates 
$\{\gamma_{\mu}, \gamma_{\nu}\} = 2 \eta_{\mu\nu}$.

\medskip

\noindent{\bf The IIA Case: $p=2k$,  $k=0,\ldots,4 $ }

We have defined a $(p+1)$-form 
\begin{equation}
\rho_{p+1}= {1 \over (p+1)!}\rho_{\mu_{1} \ldots \mu_{p+1}}
 d\sigma^{\mu_{1}} \ldots d\sigma^{\mu_{p+1}}
\end{equation}
and represented $\rho^{(p)}$ as: 
\begin{equation}
\rho^{(p)}  =  {1 \over (p+1)!}\epsilon^{\mu_{1} \ldots \mu_{p+1}}
                                   \rho_{\mu_{1} \ldots \mu_{p+1}}.
\end{equation}
For $p=2k$ it follows from eqs. (\ref{rhoandT}-\ref{SAeqn}) that
\begin{equation}
\rho_{2k+1} =   \sum_{n=0}^{k} \GE ^{(k-n+1)} \frac{{\cal F}^{n}}{n!}
\frac{\psi^{2(k-n)+1}}{(2(k-n)+1)!}.
\end{equation}
The expression can be most easily examined by choosing a canonical basis for ${\cal F}$.
The point is that both sides of the the equation
we are attempting to prove are Lorentz invariant, and a $(p+1)$-dimensional
Lorentz transformation can bring ${\cal F}$ to the special form
\begin{equation}
  {\cal F}  = \sum_{i = 1} ^ {k} \Lambda_{i} \,
                               d \sigma^{2i-1} \wedge d \sigma^{2i}. 
\end{equation}
Since $p$ is even, there is necessarily a row and a column of zeroes, which we have
chosen to associate with the time direction, thereby making ${\cal F}$ purely
magnetic. The argument works the same if there are electric components.
In this basis, defining $\gamma_{i}^{[2]} \equiv \gamma_{2i-1} \gamma_{2i}$, 
\begin{equation}
 \rho^{(2k)}= \sum_{n=0}^{k}\!\!
                 \sum_{\stackrel{i_{1}  < \ldots <i_{n}}
                                {i_{n+1}< \ldots <i_{k}}}
           \! \!  \GE^{(k-n+1)} \! \Lambda_{i_1} \cdots \Lambda_{i_{n}} 
                  \gamma_{0} 
                  \gamma_{i_{n+1}}^{[2]} \cdots \gamma_{i_{k}}^{[2]}, 
\end{equation}
where $(i_{1}, \ldots , i_{k})$ is a permutation of the numbers $(1, \ldots ,k)$.
This can be rewritten in the much more transparent form
\begin{equation}
 \rho^{(2k)}=\GE \gamma_{0} \prod_{i=1}^{k} (\Lambda_{i} +
                                                 \GE \gamma_{i}^{[2]}) \label{rho2k}.
\end{equation}
As all the $\gamma^{[2]}$'s commute with one another and with $\GE$, whereas
$\gamma_{0}$ and $\GE$ anticommute:
\begin{eqnarray}
(\rho^{(2k)})^2 &=& \! \! \prod_{i=1}^{k}(\Lambda_{i} - \GE \gamma_{i}^{[2]})
                                      \nonumber
                                    (\Lambda_{i} + \GE \gamma_{i}^{[2]})\\ 
            &=& \!\! \prod_{i=1}^{k}(1\,+ \, \Lambda_{i}^2)   .
\end{eqnarray}
Therefore,
\begin{eqnarray}
(\rho^{(2k)})^2= \!\! - {\rm det}(G+{\cal F}),
\end{eqnarray}
in this basis. This completes the proof for $p$ even.

\medskip

\noindent{\bf The IIB Case: $p=2k+1$,  $k=0,\ldots,4 $ }

The proof for $p$ odd is almost identical, and can be made very brief. 
One difference
is that ${\cal F}$ has an even number of row and columns, so the canonical form
has no rows or columns of zeroes. Thus, in canonical basis, 
${\cal F}$ contains both electric ($\Lambda_{0}$)
and magnetic ($\Lambda_{i}$, $i = 1, \ldots , k$) components
\begin{equation}
  {\cal F}  = \sum_{i = 0} ^ {k} \Lambda_{i} \,
                               d \sigma^{2i} \wedge d \sigma^{2i +1}. 
\end{equation}
This time it is convenient to define $\gamma_{i}^{[2]} \equiv \gamma_{2i} 
\gamma_{2i +1}$. Then, using eqs. (\ref{rhoB}-\ref{CBeqn}), 
one can show that the counterpart of eq. (\ref{rho2k}) is
\begin{equation}
 \rho^{(2k+1)}=  \tau_3 \tau_1 \prod_{i=0}^{k} (\Lambda_{i} -
                                                 \tau_3 \gamma_{i}^{[2]}) .
\end{equation}
The square of this also gives the desired determinant.

\section*{Appendix B: The Proof of Equation (\ref{2star})}

We wish to prove the IIA identity
\begin{equation} \label{1}
\rho^{(p)} \gamma_\mu = T^{\nu}_{(p)} ( G - {\cal F} \Gamma_{11} )_{\nu \mu}. 
\end{equation} 
The proof is the quickest in the differential form representation. 
By definitions (\ref{rhoup}-\ref{rhodown}) and (\ref{Tup}-\ref{Tdown}), 
eq.~(\ref{1}) is equivalent to:
\begin{equation}
\rho_{p+1} \gamma_{\mu} = T_p d \sigma^{\nu} 
( G - {\cal F} \Gamma_{11} )_{\nu \mu}.
\end{equation}
This allows us to combine the formulas for all even $p$'s. Using
\begin{equation} 
\rho_{A}={\displaystyle \sum_{p\, even}} 
\rho_{p+1} = e^{\cal F} S_{A}(\psi)\quad {\rm and} \quad
T_{A}={\displaystyle \sum_{p\, even}}\,\, T_{p} = e^{\cal F} C_{A}(\psi),
\end{equation}
we get
\begin{equation}\label{4}
\rho_{A} \gamma_{\mu} = T_A d \sigma^{\nu} 
( G - {\cal F} \Gamma_{11} )_{\nu \mu}.
\end{equation}
The key to the proof is the relation\footnote{This is equivalent to 
       $ \gamma_{\mu_{1}\ldots \mu_{n}} \gamma_\mu =
   n  \gamma_{[\mu_{1}\ldots \mu_{n-1}} G_{\mu_{n}]\mu} +
        \gamma_{\mu_{1}\ldots \mu_{n}\mu}. $}
\begin{equation} \label{2} 
\frac{\psi^{n}}{n!} \gamma_{\mu} \,=\,\frac{\psi^{n-1}}{(n-1)!}
d \sigma^{\nu} G_{\nu \mu} \,+\,{(-1)^{n} \over (n+1)!}
i_{e_{\mu}}(\psi^{n+1}),
\end{equation}
where $i_{e_{\mu}}$ denotes the interior product operator induced by
$e_{\mu}=\frac{\partial}{\partial \sigma^{\mu}}$. 
This is a consequence of the definition of $i_X$, 
\begin{equation}
i_X \omega = \frac{1}{n!} {\displaystyle \sum_{s=1}^n  (-1)^{s-1}
X^{\mu_s} \omega_{\mu_1 \ldots \mu_s \ldots \mu_n} 
      d\sigma^{\mu_1} \ldots d\sigma^{\mu_{s-1}} d\sigma^{\mu_{s+1}} 
                       \ldots d\sigma^{\mu_n}},
\end{equation}
for an $n$-form $\omega$ and a vector field $X$.
Using eq.~(\ref{2}), it follows directly that
\begin{equation} \label{3}
\rho_{A} \gamma_{\mu} = T_A d \sigma^{\nu} G_{\nu \mu} - e^{\cal F}
 i_{e_{\mu}}\!(C_{A}(\psi)) \Gamma_{11}. 
\end{equation}
It must be kept in mind that eq.~(\ref{3}) is a set of equations relating 
differential forms of order $p+1$, the dimension of the world volume.
As $e^{\cal F} C_{A}(\psi)$ is a $p+2$ form, and therefore vanishes, we have
\begin{equation}
 e^{\cal F} i_{e_{\mu}}\! (C_{A}(\psi)) \Gamma_{11} =
-  i_{e_{\mu}}\! (e^{\cal F}) C_{A}(\psi) \Gamma_{11} =
 -T_{A} d\sigma^{\nu} {\cal F}_{\mu \nu} \Gamma_{11} .
\end{equation}
This gives the second term on the rhs of eq.~(\ref{4}) 
and completes the proof for the IIA case.

The proof for IIB is similar, except that $ \tau_{3} $ anticommutes 
with $ S_B (\psi) $ which introduces an extra minus sign in the second term. 
This is precisely what is needed, because the IIB version of eq.~(\ref{1}) is
\begin{equation}
\rho^{(p)} \gamma_\mu = T^{\nu}_{(p)} ( G + {\cal F} \tau_3 )_{\nu \mu}. 
\end{equation}

\section*{Appendix C: The Proof of Equation (\ref{3star})}

We wish to prove the identity
\begin{equation}
d\bar\theta\delta T_p d\theta + 2 \delta \bar\theta d T_p d\theta = 0.
\end{equation}
We start with the IIA case. Summing over all even values of $p$ gives 
\begin{equation}
d\bar\theta\delta T_A d\theta + 2 \delta \bar\theta d T_A d\theta = 0,
\end{equation}
where $T_A$ is given in eqs. (\ref{rhoandT}-\ref{CAeqn}). We evaluate this using
\begin{equation}
\delta T_A = \delta [ e^{{\cal F}} C_A ( \psi ) ] = e^{{\cal F}} ( \delta {\cal F} C_A + 
\delta C_A ) ,
\end{equation}
\begin{equation}
\delta C_A = \Gamma_{11} [ \delta \psi S_A ] = 2 \delta \bar\theta \Gamma^m d 
\theta  \Gamma_{11} [ \Gamma_m S_A ] ,
\end{equation}
and
\begin{equation}
\delta {\cal F} = 2 \delta \bar\theta \Gamma_{11} \Gamma_{m_{1}} d \theta \Pi^{m_{1}},
\end{equation}
where the brackets denote antisymmetrisation of all enclosed $ \Gamma $ matrices. 
Dropping an overall factor of $ 2 e^{{\cal F}} $ and collecting the 
coefficient of $2k+1$ $ \Pi $'s, we 
can write the contribution to $ d \bar\theta \delta T_A d \theta $ as
\begin{equation}
{1 \over 2k!}\delta \bar\theta \Gamma_{11} \Gamma_{[m_{1}} d \theta
d \bar\theta \Gamma_{11}^{k+1} \Gamma_{m_{2}\ldots m_{2k+1}]} d \theta + 
{1 \over (2k+1)!}\delta \bar\theta \Gamma^m d \theta                      
d \bar\theta \Gamma_{11}^{k} \Gamma_{mm_{1}\ldots m_{2k+1}} d \theta .
\label{deltaTA}
\end{equation}

The contribution from $ 2 \delta \bar\theta d T_A d \theta $ has precisely the same form,
except that the $ \delta $'s appear in the second factor of each term. 
Both of the  terms in (\ref{deltaTA})
have the structure $\delta \bar\theta X d\theta d\bar\theta Y d \theta$,
which involves $X_{\alpha(\beta} Y_{\gamma\delta)}$. However, when 
$d \bar\theta X d\theta \delta \bar\theta Y d \theta$ is added, the totally symmetric
combination $X_{(\alpha\beta} Y_{\gamma\delta)}$ is formed. 
This implies that it is sufficient to prove the vanishing of the 
sum of the two terms above with $ \delta \theta $ replaced by $ d \theta $ 
\begin{equation}
(2k+1)d \bar\theta \Gamma_{11} \Gamma_{[m_{1}} d \theta
d \bar\theta \Gamma_{11}^{k+1} \Gamma_{m_{2}\ldots m_{2k+1}]} d \theta + 
d \bar\theta \Gamma^m d \theta                      
d \bar\theta \Gamma_{11}^{k} \Gamma_{mm_{1}\ldots m_{2k+1}} d \theta ,
\label{closure}
\end{equation}
since this enforces total symmetrization.
This is an identity we need anyway, to prove closure of the
Wess-Zumino forms $I_{p+2}$.

The next step is to transform the second term in (\ref{closure}) using the formula
\begin{equation}\label{key-eq}
\Gamma_m \Gamma_{m_{1}\ldots m_{2k+1}} = \Gamma_{mm_{1}\ldots m_{2k+1}} +
(2k+1) \eta_{m[m_{1}} \Gamma_{m_{2}\ldots m_{2k+1}]}.
\end{equation}
The two terms on the RHS of this formula have opposite symmetry in spinor indices, so only
one of them survives when sandwiched in between $ d \theta $'s. In the
present (IIA) case, it is
the first one that survives, which means that we can pull out a factor $ \Gamma_m $ 
from the antisymmetrized product for free.
Next, eq. (\ref{3spinors}) for Majorana-Weyl spinors in 10 dimensions 
implies that
\begin{equation}
d \bar\theta \Gamma^m d \theta    
d \bar\theta \Gamma_{11} \Gamma_m\Gamma_{m_{1}\ldots m_{2k+1}} 
\Gamma_{11}^{k+1} d \theta = 
- d \bar\theta \Gamma_{11} \Gamma^m d \theta    
d \bar\theta \Gamma_m\Gamma_{m_{1}\ldots m_{2k+1}} 
\Gamma_{11}^{k+1} d \theta.
\label{lastofC}
\end{equation}
We now use eq.~(\ref{key-eq}) a second time. This time only the second
term on the RHS survives, because we have removed a $ \Gamma_{11} $,
which reverses the symmetry. This leaves the negative of the first term
in eq. (\ref{closure}), and thus the proof is
complete. 

The IIB proof is essentially the same. 


\medskip

\section*{Appendix D: Verification of the Supersymmetry}

We wish to verify invariance of the action 
\begin{equation}
S^{(p)} = - \int \sqrt{-
{\rm det}\, M^{(p)}}\,  d^{p+1} \sigma,
\end{equation} 
where
\begin{equation}
M_{\mu\nu}^{(p)} = \eta_{\mu\nu} + F_{\mu\nu} + \partial_\mu \phi^i
\partial_\nu \phi^i - 2 \bar\lambda (\Gamma_\mu + \Gamma_i \partial_\mu \phi^i)
\partial_\nu \lambda
+ \bar\lambda \Gamma^m \partial_\mu \lambda \bar\lambda \Gamma_m \partial_\nu
\lambda,
\end{equation}
under the global supersymmetry transformations.  These transformations are given by
$\Delta$ variations, where  $\Delta = \Delta' + \delta_\xi$,
\begin{eqnarray}
\Delta' \lambda &=& \bar\epsilon_1 + \bar\epsilon_2 \zeta^{(p)}\nonumber \\
\Delta' \phi^i &=& (\bar\epsilon_1 - \bar\epsilon_2 \zeta^{(p)}) \Gamma^i
\lambda \nonumber \\
\Delta' A_\mu &=& (\bar\epsilon_2 \zeta^{(p)} - \bar\epsilon_1) (\Gamma_\mu +
\Gamma_i \partial_\mu \phi^i)\lambda
+ \left({1\over 3} \bar\epsilon_1 - \bar\epsilon_2 \zeta^{(p)}\right)
\Gamma_m \lambda \bar\lambda \Gamma^m \partial_\mu \lambda
\end{eqnarray}
and $\delta_\xi$ represents a general coordinate transformation with parameter
\begin{equation}
\xi^\mu = (\bar\epsilon_2 \zeta^{(p)} - \bar\epsilon_1) \Gamma^\mu \lambda.
\end{equation}

The matrix $M_{\mu\nu}^{(p)}$ does not transform under $\delta_\xi$ as a tensor,
because $\eta_{\mu\nu}$ is not a tensor and $\Gamma_\mu$ is not a vector.
However, it is easy to compensate for this, obtaining
\begin{equation}
\delta_\xi M_{\mu\nu}^{(p)} = \xi^\rho \partial_\rho M_{\mu\nu}^{(p)} +
\partial_\mu \xi^\rho M_{\rho\nu}^{(p)} + \partial_\nu \xi^\rho M_{\mu\rho}^{(p)}
- \partial_\mu \xi_\nu - \partial_\nu \xi_\nu + 2 \partial_\mu \xi^\rho
\bar\lambda \Gamma_\rho \partial_\nu \lambda.
\end{equation}
The first three terms gives a variation of $\sqrt{-{\rm det}\,
M^{(p)}}$ equal to $\partial_\mu (\xi^\mu \sqrt{-{\rm det}\,M^{(p)}})$, as usual for a
scalar density in relativity.  Thus it remains to consider the effective
transformation of $M_{\mu\nu}^{(p)}$:
\begin{equation}
\Delta' M_{\mu\nu}^{(p)} - \partial_\mu \xi_\nu - \partial_\nu \xi_\mu + 2
\partial_\mu \xi^\rho \bar\lambda \Gamma_\rho \partial_\nu \lambda.
\end{equation}
A somewhat lengthy, but straightforward, calculation shows that this equals
\begin{equation}
- 4 \bar\epsilon_2 \zeta^{(p)} \gamma_\mu \partial_\nu \lambda.
\end{equation}
The consequences of this term now need to be analyzed.

The contribution of the term given above to $\Delta S^{(p)}$ is
\begin{equation}
4 \int \sqrt{-{\rm det}\, M^{(p)}} \{(M^{(p)})^{-1}\}^{\nu\mu} \bar\epsilon_2
\zeta^{(p)} \gamma_\mu \partial_\nu \lambda\, d^{p+1}\sigma.
\end{equation}
To simplify this we specialize the covariant identity in eq. (\ref{referinD}) to the static
gauge under consideration.  Doing this gives
\begin{equation}
\sqrt{-{\rm det}\, M^{(p)}} \{(M^{(p)})^{-1}\}^{\nu\mu} \gamma_\mu =
\tilde{\zeta}^{(p)} t_{(p)}^\nu,
\end{equation}
where $t_{(p)}^\nu$ denotes the $(11)$ block of $T_{(p)}^\nu$ in the IIA case
and the $(12)$ block of $T_{(p)}^\nu$ in the IIB case.  Using eqs. (\ref{TA2by2})
and (\ref{TB2by2}), one sees that $t_{(p)}^\nu$ is determined by the $p$-form
part of ${\rm exp}\,({\cal F} +\psi)$ in the same way that $T_p$ determines
$T_{(p)}^\nu$.

Using the relation $\zeta^{(p)} \tilde{\zeta}^{(p)} = 1$, we are left with
\begin{equation}
4 \int \bar\epsilon_2 t_{(p)}^\nu \partial_\nu \lambda\, d^{p+1} \sigma.
\end{equation}
However, as explained in sect. (4.2),
\begin{equation}
t_{(p)}^\nu\partial_\nu \lambda = \partial_\nu J_{(p)}^\nu,
\end{equation}
where $J_{(p)}^\nu$ is the supercurrent.  Therefore, altogether,
\begin{equation}
\Delta S^{(p)} = \int \partial_\mu \big( - \xi^\mu \sqrt{-{\rm det}M^{(p)}} + 2
\bar\epsilon_2  J_{(p)}^\mu \big) d^{p+1} \sigma,
\end{equation}
which is the integral of a total derivative and vanishes for
suitable asymptotic boundary conditions.

In retrospect, the result derived here should have been obvious in the first place.
The supersymmetry variations of the gauge-fixed theory correspond to
a combination of supersymmetry variations and compensating gauge 
transformations in the gauge-invariant theory. The term $S_1$
of the gauge-invariant action is invariant under the global
supersymmetry transformations, and the result that we have found just
corresponds to the contributions of the compensating gauge transformations,
evaluated in the static gauge.

\medskip
\section*{Appendix E: Dimensional Reduction of $\zeta^{(p)}$}

As indicated in eq. (\ref{gamma2by2}), 
$\zeta^{(p)}$ is the (12) block of $\gamma^{(p)}$.
Moreover $\gamma^{(p)}$ is related to $\rho^{(p)}$ by $\gamma^{(p)} = (-{\rm det}\,
M^{(p)})^{-1/2} \rho^{(p)}$.  We showed in sect. (4.3) that, dropping the
dependence on one coordinate ($\sigma^p$, say), $M^{(p)}$ reduces to
$M^{(p-1)}$.  Therefore, we need to study the dimensional reduction of the (12)
block of $\rho^{(p)}$.  However, $\rho^{(p)}$ is conveniently described by
a $(p + 1)$-form, as explained in sect. (3.2), and these are conveniently
summed to give $\rho_A$ and $\rho_B$.  Their (12) blocks, 
given in eqs. (\ref{rhoA2by2})
and (\ref{rhoB2by2}), can be combined to give
\begin{equation}
\rho = e^{{\cal F}+\psi}.
\end{equation}
The $(p + 1)$-form part of this, evaluated in the static gauge, determines
$\zeta^{(p)}$ for all $p$.

Now we evaluate ${\cal F}^{(p)} = F^{(p)} - b^{(p)}$ and $\psi^{(p)}$ in the
static gauge.  With the conventions described in the text, one obtains in both
the IIA and IIB cases
\begin{eqnarray}
b^{(p)} &=& \bar\lambda \Gamma_{\rho} d\lambda d\sigma^{\rho}
 + \bar\lambda \Gamma_i
d\lambda d\phi^i\nonumber \\
\psi^{(p)} &=& \Gamma_{\rho} d\sigma^{\rho}
 + \Gamma_i d\phi^i + \Gamma_m \bar\lambda
\Gamma^m d\lambda,
\end{eqnarray}
where $\rho$ runs from $0$ to $p$, $i$ runs from $p + 1$ to $9$, and $m$ runs
from $0$ to $9$.  Upon dimensional reduction, dropping $\sigma^p$ derivatives,
\begin{eqnarray}
{\cal F}^{(p)} &\rightarrow & {\cal F}^{(p-1)} + \bar\lambda \Gamma_p d\lambda
d\phi^p - (d \phi^p + \bar\lambda \Gamma_p d\lambda) d\sigma^p \nonumber \\
\psi^{(p)} &\rightarrow & \psi^{(p-1)} - \Gamma_p d\phi^p + \Gamma_p d\sigma^p.
\end{eqnarray}
The next step is to see what these imply for $\exp {\cal F}^{(p)}$ and $\exp
\psi^{(p)}$.  Since the extra terms in the reduction of ${\cal F}^{(p)}$ square
to zero,
\begin{equation}
e^{{\cal F}^{(p)}} \rightarrow \big(1 + \bar\lambda \Gamma_p d \lambda d \phi^p -
(d \phi^p +  \bar\lambda \Gamma_p d\lambda)d \sigma^p\big) e^{{\cal F}^{(p-1)}}.
\end{equation}
The extra terms in $\psi^{(p)}$ also square to zero, but care is required since
they do not commute with $\psi^{(p-1)}$.  The general formula that applies to
such a case is that the part of $\exp (X + Y)$ that is linear in $Y$ is given
by $(Y + {1\over 2} [X, Y] + {1\over 6} [X, [X,Y]] + \ldots )e^X$.  Using this,
we find that
\begin{equation}
e^{\psi^{(p)}} \rightarrow (1 + (\Gamma^p + d\phi^p + \bar\lambda \Gamma^p
d\lambda) d\sigma^p - (\Gamma^p + \bar\lambda \Gamma^p d \lambda) d\phi^p)
e^{\psi^{(p - 1)}}.
\end{equation}

We now require the part of the dimensional reduction of $\exp ({\cal F}^{(p)} +
\psi^{(p)})$ that is proportional to $d\sigma^p$.  Several terms cancel and we
obtain
\begin{equation}
e^{{\cal F}^{(p)} + \psi^{(p)}} \rightarrow \Gamma_p d\sigma^p e^{{\cal
F}^{(p-1)} + \psi^{(p-1)}} = (-1)^p \Gamma_p e^{{\cal F}^{(p - 1)} +
\psi^{(p-1)}} d \sigma^p.
\end{equation}
This result implies that
\begin{equation}
\zeta^{(p)} \rightarrow (-1)^p \Gamma_p \zeta^{(p - 1)},
\end{equation}
as asserted in the text.

\end{document}